\documentclass{iaus}
\title{Galactic and Extragalactic Distance Scales:\\
The Variable Star Project}
\author{Michael W. Feast}
\affiliation{Astronomy Department, University of Cape Town}
\begin{document}
\maketitle
\begin{abstract}
 This paper summaries the status of a large project to
improve distance scales of various classes of variable stars. This  
is being carried out by a large group in Cape Town, Japan, England and the
USA. The results are illustrated by giving the distances
to the Large Magellanic Cloud and the Galactic Centre ($R_{o}$) as well as the
value of the Hubble Constant, $H_{o}$, based on our current results.
The classes of variables considered are; Classical Cepheids, Type II Cepheids,
RR Lyrae stars, O- and C- type Miras.
\end{abstract}
\section{Introduction}
In studying the structure and dynamics of our own Galaxy as well as the
distances and distributions of galaxies in the nearby Universe we need accurate
distance tracers. These tracers should be relatively common, easily
recognizable, intrinsically luminous and of well defined absolute magnitudes.
Certain classes of variable stars fit all these criteria. It is well known,
for instance that 
the brightnesses of classical Cepheids in nearby galaxies are used to calibrate
the luminosities of supernovae of type Ia, and that this leads to a
determination of the Hubble constant ($H_{o}$). But other classes of
variable stars are becoming of increasing importance in setting distance
scales. This is not only because they provide, or should provide, a
confirmation of the classical Cepheid scale when applied to the
same object (e.g. the LMC). There are many systems in which classical
Cepheids are not present (e.g. globular clusters, the Galactic bulge,
dwarf spheroidal galaxies etc.) and for which other classes of variable
stars can provide distances.
In addition, 
a knowledge of the luminosities
of variable stars is essential if we are to understand their structure and
evolution.

A group of us have been interested for some years in the calibration
of the luminosities of variable stars.  
The present paper is a short review of the 
current status of this work.
The principal workers in Cape Town, 
in recent years, have been; Patricia Whitelock,
Dave Laney, John Menzies and the writer; In Japan, Noriyuki Matsunaga and
his many colleagues; In England, Floor van Leeuwen; and in the United States,
Fritz Benedict and a large group of colleagues as well as Tom Kinman.
A full list of co-workers will be found in the references at the end of
this paper.

\section{The Variable Stars}
 Our work has focused on five main classes of variable stars:\\
The classical Cepheids, are relatively young objects;\\   
The type II Cepheids have periods in the same range as the classical
Cepheids but are old stars of low mass, some of which are found in
globular clusters;\\ 
The RR Lyrae variables are 
short period (less than a day)
variables, many of which, too, are found in globular clusters;\\ 
Miras are large amplitude,
long period, variables at the tip of the AGB and come in 
two main classes, oxygen-rich (O-Miras) and carbon-rich (C-Mira).

In some of these classes, the variables were known to follow period-luminosity
(PL) relations. In others, we have established such relations.
We can write in each case;
\begin{equation}
M_{x} = \alpha \log P + \beta
\end{equation}
where $M_{x}$ is the absolute magnitude at some wavelength, $x$, 
$P$ is the pulsation period in days, and 
$\alpha$ and $\beta$ are constants to be determined. To illustrate 
the results of our work in determining $\alpha$ and $\beta$,  I will
give values for the Hubble Constant, the distance modulus of
the Large Magellanic Cloud (LMC) and the distance from the Sun to
the Centre of our Galaxy ($R_{o}$). 

In the case of the RR Lyrae variables
the relation;
\begin{equation}
M_{V} = a[Fe/H] + b
\end{equation} is often used.

\section{The Classical Cepheids}
 The Cape Town group has been involved in two 
recent projects on the measurement
of the 
trigonometrical parallaxes of nearby classical Cepheids and their use
in defining PL relations at various wavelengths. The first project
(Benedict et al. 2007) combined ground-based and HST data to derive parallaxes.
In the second project (van Leeuwen et al. 2007) the new revision of the 
Hipparcos
parallaxes (van Leeuwen 2007) was used. Our final results combine these two
data sets. We used these new trigonometrical parallaxes to obtain PL
relations both in the optical region and in the near infrared.
Perhaps the most interesting results were obtained for $M_{W}$, the
``reddening-free" parameter given by:
\begin{equation}
 W = V -2.45(V-I).
\end{equation}
In that case we find for the coefficients of equation 2.1,
$\alpha = -3.29 \pm 0.15$ and $\beta = -2.58 \pm 0.03$.  
These values  can now be used to redetermine the distances of the
galaxies which were used by Sandage et al. (2006) to calibrate the
luminosity of 
type Ia supernovae and from that to derive the Hubble constant. 
Doing this we find $H_{o}= 70 \pm 5\, \rm km s^{-1} Mpc^{-1}$ 
in good agreement with the WMAP
value of 
$73 \pm 3\, \rm km s^{-1} Mpc^{-1}$ 
assuming a $\Lambda CDM$ model (Spergel et al. 2007).
In general one has to take into account in such a derivation from
classical Cepheids the possible effect of any differences in metallicity
between the calibrating Cepheids and those in the extragalactic systems
being studied. Fortunately in the case of the SNIa calibration the
average metallicity of the Cepheids in question are believed to be
close to that of Cepheids in the solar neighbourhood and any metallicity
correction is thus negligible.

In principle both the slope and zero-point of the PL relation could
depend on the metallicity of the Cepheids. A test can be made 
of a possible slope dependence
using LMC Cepheids since they are metal deficient
compared with those in the solar neighbourhood. In the case of the
relation in $M_{W}$, observations of LMC Cepheids by the OGLE group 
(Udalski et al. 1999))
shows that $\alpha = -3.29 \pm 0.01$ (see Benedict et al. 2007). Thus
within the limits of their uncertainties, the values of $\alpha$
are the same in the Galaxy as in the LMC, contrary to some suggestions.
However, observations of Cepheids in extragalactic systems in
which they have a range of metallicities show that the zero-point,
$\beta$, in the PL($M_{W}$) relation does depend on metallicity and this
needs to be taken into account in deriving the distance modulus of the LMC.
Without a metallicity correction the PL($M_{W}$) relation gives a modulus of 
$18.52 \pm 0.03$. Using a metallicity
correction derived by Macri et al. (2006) from their work on Cepheids in the
galaxy NGC4258 leads to a corrected modulus of $18.39 \pm 0.05$, somewhat
smaller than the conventional, but poorly defined, value of 18.50.

Some years ago (Feast \& Whitelock 1997) the Hipparcos proper motions of
Galactic Cepheids were used to derive the Oort constants of differential
galactic rotation. The results were;  
$A = 14.82 \pm 0.84\, {\rm km\,s^{-1}kpc^{-1}}$  and
$B = -12.37 \pm 0.64\, {\rm km\,s^{-1}kpc^{-1}}$. These values are essentially
independent of the adopted distance scale. The results were also
combined with radial velocity observations (Pont et al. 1994) to 
estimate the distance to the Galactic Centre ($R_{o}$). This result
does depend on the adopted scale. Thus as a result of our revision of the
Cepheid scale the value of $R_{o}$ will change. The 1997 value was 
$8.5 \pm 0.5 \,\rm kpc$. It seems likely that the new scale will result in a
reduction of this value, possibly to $\sim 7.5\, \rm kpc$, but a full 
re-reduction
of the available data has not yet been made.

There are other ways besides trigonometrical parallaxes of determining
the luminosities of Cepheids (e.g. pulsation parallaxes, membership of
open clusters). However, the trigonometrical parallax method is the
most direct and subject to the fewest assumptions. It has recently
been suggested that a distance to the Cepheid RS Pup can be obtained with
exceptionally high accuracy from observations of its surrounding nebula 
(Kervella
et al. 2008). However this depends on the adoption of a model for the
nebulosity for which there is no a priori justification.
On the other hand, it is possible
to derive a nebula model which is consistent with some adopted distance.
If we adopt a distance base on equation 2.1 with the constants given
above, it can be shown (Feast 2008) that 
the data fit a model
in which the nebulosity of RS Pup is confined to an equatorial disc at
a small angle to the plane of the sky. This is a very satisfactory model 
from an astrophysical point of view.

\section{The Type II Cepheids}
 PL relations at optical wavelengths for type II Cepheids have been
suggested from time to time, but the scatter in them is quite large.
However on the basis of a large scale survey for variables in
globular clusters in the near infrared using the Japanese-South African 
1.4m telescope,
Matsunaga et al. (2006) showed that the type II Cepheids in these clusters
fell on a well defined, narrow, PL($M_{K}$) relation. $K$ is the magnitude  
at $2.2\,\rm m \mu$. 
The fact that some globular clusters contain type II Cepheids with a range of 
periods and also that
the relative distances of globular clusters is reasonably
well established means  
that the slope of equation 2.1, with $M_{x} \equiv M_{K}$,
may be taken from the Matsunaga et al. paper, i.e. $\alpha = -2.41 \pm 0.05$.
However, the absolute distance scale for globular clusters is still 
somewhat uncertain.  It is desirable that an independent way is found
to calibrate the type II Cepheid PL($M_{K}$) relation. Some progress in this
direction has been made by our group (Feast et al. 2008) using both
trigonometrical and pulsational parallaxes of a few stars in this class.
In this case the pulsational parallaxes are the most useful, provided
the assumptions made in deriving them are correct. The most crucial
assumption is the value of the projection factor used to convert measured
radial velocities to the pulsational velocities of the star. In this work
the projection factor was based on values derived from pulsation work
on classical Cepheids with good trigonometrical parallaxes. 
The pulsation parallaxes of the two type II Cepheids, SW Tau and V553 Cen
show that they deviate by only 0.02 mag in the mean from a PL($M_{K}$)
relation based on the globular cluster slope and with $\beta = -1.02$.
The globular clusters work suggests that the PL($M_{K}$) relation for these
stars has little if any dependence on metallicity. Having derived this
zero-point we can apply it to observations of type II Cepheids in the LMC
(Alcock et al. 1998).
We then find an LMC modulus of $18.37 \pm 0.09$
in good agreement with the Classical Cepheid modulus discussed above.
Applying our PL({$K$}) relation to type II Cepheids in the
Galactic Bulge ( Groenewegen et al. 2008) leads to 
a distance to the Galactic Centre of $R_{o} = 7.64 \pm 0.21$kpc.
This is in excellent agreement with  the value ($7.73 \pm 0.32$kpc)
derived from the motion of a star round the central blackhole
(Eisenhauer et al. 2005, Zucker et al. 2006).
However it would obviously be desirable to obtain good distances of
more type II Cepheids. A current uncertainty is that there is a
discrepancy between the trigonometrical and pulsational parallaxes of
the type II Cepheid $\kappa$ Pav. This may be due to it being a close binary
for which there is some evidence. But further work is clearly required.
Work is 
in fact in progress, led by Fritz Benedict, and including other members of
our group, to obtain HST trigonometrical parallaxes of two  type II
Cepheids, one of which is $\kappa$ Pav.

\section{The RR Lyrae Variables}
  The RR Lyrae distance scale was also discussed by Feast et al. (2008).  
The luminosities of RR Lyrae variables are frequently analysed on the
basis of equation 2.2 above. There has much debate on the proper value of
$a$ to use in this equation.
We adopted 0.214 from  a study of LMC RR Lyraes (Gratton et al. 2004). 
There is a good HST-based trigonometrical parallax of RR Lyrae itself
(Benedict et al. 2002).
Taking a mean of this with the revised Hipparcos parallax leads to 
an absolute magnitude of $M_{V} = +0.54 \pm 0.11$. Unfortunately
there are no other RR Lyraes with individually useful trigonometrical
parallaxes. This is particularly unfortunate since Catelan \& Cort\'{e}s
(2008) predict on the basis of Str\"{o}mgren photometry that RR Lyrae itself
is overluminous for its metallicity compared with average members
of the RR Lyrae type. Indeed we find that this star is $0.16 \pm 0.12$mag
brighter in $V$ that would be predicted from equation 2.2 and the LMC
RR Lyraes taken at a modulus of 18.39 as given by the classical Cepheids
(see section 3).  
This agrees within the errors with
the overluminosity predicted
by Catelan \& Cort\'{e}s ($0.06 \pm 0.01$mag). 
Again, work in progress with the HST should lead to
new trigonometrical parallaxes of other RR Lyrae stars.

There is however an additional problem with using equation 2.2 for RR Lyraes
in that the coefficient, $a$, appears to depend on environment. For instance,
Clementini et al. (2005) find a different value for the Sculptor galaxy
from that in the LMC. It may be better in the future to use an RR Lyrae
PL($M_{K}$) relation (Longmore, Fernley \& Jameson 1986, Sollima, Cacciari \& 
Valenti 2006).

\section{The Mira Variables}
  PL($M_{K}$) and  PL($M_{bol}$) relations for O- and C-Mira 
variables in the LMC
were given by Feast et al. (1989). The slope of the PL({$M_{K}$) 
relation for the O-Miras
has now been slightly revised using additional data (Whitelock, Feast
\& van Leeuwen 2008). The zero-point of this relation was also revised
by combining revised Hipparcos parallaxes, VLBI observations and 
distances from Miras
in globular clusters.  If this data is combined with the work on O-Miras in the
LMC it leads to a LMC modulus of $18.49\pm 0.07$. This result is not in
conflict with the somewhat smaller moduli derived from classical and type II
Cepheids above. Further work on the distance scales of globular clusters
and more VLBI work should lead to improvements in the O-Mira 
PL($M_{K}$) relation.

Whitelock, Feast \& van Leeuwen (2008) also discussed the PL($M_{K}$) relation
for C-Miras  based on trigonometrical parallaxes
though the uncertainties are greater in this case. Somewhat
earlier, extensive infrared photometry and radial velocity work on
Galactic C-Miras was carried out (Whitelock, Feast, Marang \& Groenewegen 2006,
Menzies, Feast \& Whitelock 2006, Feast, Whitelock \& Menzies 2006). 
Adopting the value of the Oort constant $A$ as derived from classical
Cepheids (see section 3) it was possible to combine the radial velocity,
period and photometric data to derive a PL($M_{K}$) zero-point. Putting
this result together with the result from the parallaxes, just mentioned,
leads to a PL($M_{K}$) zero-point which together with data on C-Miras
in the LMC gives an LMC modulus of $18.31 \pm 0.20$. This agrees satisfactorily
with the Classical and type II Cepheid LMC moduli. However the uncertainty
is rather large. It would seem best for the present to rely for a C-Mira
zero-point on a value taken from the LMC with a Cepheid based distance.
An example of the current interest in C-Miras is their use
as both distances tracers and indicators
of population age in the dwarf galaxies of the local group (see for
instance, Menzies et al. (2002), Menzies et al. (2008), Whitelock et al.
(in preparation)).

\section{Conclusions}
  The above discussion shows that the distance scales we obtain for
the various classes of variable stars are quite consistent with one another.
Of particular interest is that this scale leads to a modulus of the
LMC somewhat smaller than conventionally assumed. The results also
lead to a distance to the Galactic Centre which is also smaller than that
adopted in recent times but it agrees well with the value derived
from the motion of a star round the central blackhole.
 Work is in progress to further strengthen these distance scales.
\section{Acknowledgments}
   As indicated in the Introduction, this paper summarizes the work of
many people in various parts of the world.

\end{document}